\documentclass[11pt,a4paper,english]{article}
\pdfoutput=1 

\usepackage{jheppub}
\usepackage{color}
\usepackage{babel}
\usepackage{float}
\usepackage{amssymb}
\usepackage{graphicx}
\usepackage{esint}

\makeatletter

\@ifundefined{textcolor}{}
{%
 \definecolor{BLACK}{gray}{0}
 \definecolor{WHITE}{gray}{1}
 \definecolor{RED}{rgb}{1,0,0}
 \definecolor{GREEN}{rgb}{0,1,0}
 \definecolor{BLUE}{rgb}{0,0,1}
 \definecolor{CYAN}{cmyk}{1,0,0,0}
 \definecolor{MAGENTA}{cmyk}{0,1,0,0}
 \definecolor{YELLOW}{cmyk}{0,0,1,0}
}

\@ifundefined{date}{}{\date{}}
\@ifundefined{showcaptionsetup}{}{%
 \PassOptionsToPackage{caption=false}{subfig}}
\usepackage{subfig}
\makeatother

\def\bea{\begin{eqnarray}}
\def\eea{\end{eqnarray}}

\title{A Simplified Model for Dark Matter Interacting Primarily with Gluons}

\author[a]{Rohini M. Godbole,}
\author[a,b]{Gaurav Mendiratta}
\author[b]{and Tim M.P. Tait}

\affiliation[a]{Center for High Energy Physics,\\ Indian Institute of Science,\\ Bangalore, India 560012}
\affiliation[b]{Department of Physics and Astronomy,\\ University of California,\\Irvine, CA, USA 92697}

\emailAdd{rohini@cts.iisc.ernet.in}
\emailAdd{gaurav@cts.iisc.ernet.in}
\emailAdd{ttait@uci.edu}

\abstract{We consider a simple renormalizable model providing a UV completion for dark matter whose interactions with the
Standard Model are primarily via the gluons.  The model consists of scalar dark matter interacting with scalar
colored mediator particles.  A novel feature is the fact that 
(in contrast to more typical models containing dark matter whose interactions
are mediated via colored scalars) the colored scalars typically decay into multi-quark final
states, with no associated missing energy.  We construct this class of models and examine
associated phenomena related to dark matter annihilation, scattering with nuclei, and production at
colliders.}
\preprint{UCI-HEP-TR-2015-04}
\arxivnumber{1506.01408}

\begin{document}

\maketitle
\flushbottom

\section{Introduction}

Evidence from astrophysics and cosmology points overwhelmingly to the fact 
that the Universe contains a large quantity of
non-relativistic matter that is at most weakly interacting \cite{Bertone:2004pz}.
Determining the identity of this dark matter (DM) is of paramount importance to our understanding
of particle physics, and may offer clues as to the organizing principles that underpin
the Standard Model (SM) of particle physics and its extensions.

There are two broad strategies to make progress in understanding the nature of dark matter.
Specific theories such as supersymmetric extensions of the SM 
often contain candidate particles which could play the role of dark matter, despite their
primary motivation arising from the desire to explain other mysteries .  The study
of such theories has the virtue that it
represents exploration of the vision for dark matter contained within
our best guesses for physics beyond the Standard Model.  Another line of
exploration seeks the more modest goal of describing the properties of dark matter
while remaining agnostic about its connection to more fundamental questions.  The
study of such ``simplified models" \cite{Alves:2011wf} thus
sets out to do less but has the feature that it covers a wider range of the possible
theory space for models of dark matter.  Specific modules containing
the dark matter and a mediating particle which lead to dark matter coupling
to quarks \cite{Gershtein:2008bf,Bai:2010hh,Chang:2013oia,Bai:2013iqa,An:2013xka,DiFranzo:2013vra,Garny:2014waa,Papucci:2014iwa,Harris:2014hga,Abdallah:2014hon,Malik:2014ggr}
and leptons \cite{Chang:2014tea,Bai:2014osa,Garny:2015wea}
have been constructed and investigated, as well as the universal effective field theory
limit that results when the mediating particles are very heavy compared to the energies
of interest \cite{Harnik:2008uu,Cao:2009uw,Beltran:2008xg,Beltran:2010ww,Goodman:2010yf,Goodman:2010ku,Bai:2010hh}.

Models where the interactions with gluons dominate are slightly more subtle.  $SU(3)_C$ gauge invariance
demands that couplings of gluons to the dark sector are effectively non-renormalizable interactions mediated
by loops of colored particles.  
Nonetheless, such models are interesting and provide a blueprint for theories which are best tested in direct searches
and high energy hadron colliders while largely escaping from indirect searches.  They are a worthwhile corner
of dark matter model-space to explore.
Existing proposals for simplified models with this feature can be classified in three different classes: a) the mediating particle is a vector which interacts with the SM particles via effective vertices \cite{Dudas:2013sia}, b) the mediator  is a (pseudo-)scalar which may or may not mix with the SM 
Higgs \cite{Baek:2011aa,Cotta:2013jna,Baek:2014jga,Abdullah:2014lla,Buckley:2014fba} and the dark matter is a SM singlet, c) mediator is the Higgs itself, the dark matter being
charged under the electroweak $SU(2) \times U(1)$ \cite{Cohen:2011ec,Dedes:2014hga}.  While all the options are theoretically well motivated, the options where the mediator is a  pseudo-scalar or a Higgs, both capture features found in the minimal supersymmetric standard model.

An alternative possibility presents itself when the dark matter is a scalar particle.  In that case, a quartic
interaction with any other scalar in the theory is always gauge invariant and will not mediate
dark matter decay.  If the second (mediator) scalar is colored, it will induce coupling between
a pair of dark matter scalars and gluons at one loop, represented as the dimension six interaction,
\bea
\frac{\alpha_s}{M^2} ~ |\chi|^2 ~ G^{a \mu \nu} G^{a}_{\mu \nu}
\eea
(also known as C5 \cite{Goodman:2010ku}) 
when the mediating colored scalar is heavy compared to energies of interest.  In this article, we 
explore this simplified model, examining the rich collider and astrophysical signatures of
such a construction for a variety of color representations of the scalar mediator.

Our work is organized as follows.
In section \ref{sec:Lagrangian}, we present the Lagrangian and 
couplings of the dark matter and mediator to each other and with the SM. 
In section \ref{sec:Relic-Density}
we compute the annihilation cross section and
find the range of parameters for which the dark matter would
saturate the observed density of dark matter in the Universe as a thermal
relic.
In section \ref{sec:Direct-detection}, we find the constraints from
direct detection searches for dark matter and 
in section \ref{sec:Collider-Constraints} discuss the constraints from collider searches.
We conclude in section \ref{sec:Conclusions}.

\section{Simplified Model\label{sec:Lagrangian}}

The basic module consists of a massive scalar (assumed complex for simplicity, though the modification
to a real field is simple) $\chi$ that is a gauge singlet to play the role of dark matter,
and a set of massive (typically complex) colored scalars $\phi$ 
(in representation $r$ of $SU(3)_C$) to act as the
mediator with the SM.  These basic pieces are described by the Lagrangian,
\bea
{\cal L} \supset 
\partial_{\mu}\chi^{*} \partial^{\mu}\chi - m_\chi^2 | \chi |^2
+ (D_{\mu} \phi)^{\dagger} D^{\mu}\phi - m_\phi^2 | \phi |^2
\eea
where $D_\mu \phi$ is a covariant derivative that includes interactions with the electroweak gauge
fields (in cases where $\phi$ is charged under $SU(2) \times U(1)$) and coupling to the
gluons $G^a_\mu$:
\bea
D_{\mu}\phi & \equiv &
\partial_{\mu} \phi - i g_{s} \frac{\lambda^a_r}{2} G^a_{\mu} \phi + ~{\rm Electroweak}
\eea
where $g_s$ is the strong coupling and $\lambda^a_r / 2$ are the generators of $SU(3)_C$
in representation $r$.  As discussed below, for specific color representations $r$
it is well-motivated to consider a set of fields $\phi_i$ labeled by a flavor index $i$.

The dark matter interacts with the mediator via a quartic interaction,
\bea
\lambda_d ~ |\chi |^2 |\phi|^2~.
\eea
It is interesting to note that this interaction allows for $\chi$ to be charged under a $Z_2$ symmetry forcing it to be stable,
though $\phi$ can either be $Z_2$-charged or not.  This feature, and the freedom to
choose the color representation of $\phi$ somewhat arbitrarily
result in drastically different phenomenology compared to the
usual simplified models in which dark matter interacts with a colored mediator \cite{Chang:2013oia,Bai:2013iqa,An:2013xka,DiFranzo:2013vra}.
The symmetries also permit additional quartic interactions such as,
\bea
\lambda_{\chi h} |H|^2 |\chi|^2 + \lambda_{s h} |H|^2 |\phi|^2~,
\eea
where $H$ is the SM Higgs doublet.
The former leads to Higgs portal couplings for the dark matter \cite{Burgess:2000yq}, and has been well-explored in the literature.  
While such a coupling is inevitable, we assume that its effects are sub-dominant 
to $\lambda_d$ for the purposes of our discussion.  The latter term will shift the $\phi$ mass after electroweak symmetry-breaking,
and induces couplings between the $\phi$ and the Higgs boson.  Such a coupling can be constrained by the shift it induces
in the effective coupling of the SM Higgs boson to gluons \cite{ArnoldWise:2009scalarMFV}.  In principle it is also
expected to be generically present, 
but we shall consider the case where it is also small and thus unimportant.

\subsection{Interactions of the Mediator with Quarks}

In general, the mediators can interact with quarks, allowing them to decay into hadrons.
That such decays happen is important to insure that a primordial population of $\phi$ do not
(being colored) bind with nuclei, which would be subject to strong constraints from searches
for anomalously heavy isotopes \cite{Yamagata:1993jq}.  For $SU(3)_C$ representations
$r=3, 6, 8$, interactions with a pair of quarks/anti-quarks are permitted at the
renormalizable level, provided the $SU(2) \times U(1)$ charges of $\phi$ are also
chosen appropriately.  The possibilities were tabulated in Ref.~\cite{ArnoldWise:2009scalarMFV},
which assigns flavor indices to the $\phi$ fields
such that the coupling to quarks can be governed by the principle of
minimal flavor violation (MFV) \cite{Buras:2000dm,D'Ambrosio:2002ex}.
MFV dictates that all breaking of the $SU(3)_{Q_L} \times SU(3)_{u_R} \times SU(3)_{d_R}$ flavor symmetries of the (massless) SM
be proportional to the SM Yukawa interaction matrices $Y_u$ and $Y_d$,
and is motivated to control what would otherwise be very large contributions to flavor-violating observables in
the quark sector which would be in conflict with precision measurements.

For higher color representations, coupling to quarks is still permitted but must be represented
as non-renormalizable interactions, implying the existence of additional heavy particles.

We concentrate on the specific case of a color triplet $\phi$ which has electroweak quantum 
numbers
such that it can interact with a pair of right-handed up-type quarks, though we will comment on other cases
where appropriate.  A $\phi$ that is a color triplet with charge $-4/3$ can couple to $u_i u_j$ provided
that the color indices are contracted anti-symmetrically.  MFV is implemented by choosing $\phi$ to have its
own $SU(3)_{u_R}$ flavor index, and a flavor singlet
is constructed by contracting the flavor indices anti-symmetrically,
$\epsilon_{ijk} \phi_i u_j u_k$.
This type of scalar ``diquark" bears some resemblance to the squarks of an $R$-parity-violating
supersymmetric theory.  However, their weak charges and the flavor structure of their couplings are
distinct from the supersymmetric case.

Consistently with MFV, the large top Yukawa coupling allows for deviations of coupling of 
$\phi_3$ from $\phi_{1,2}$.  If one neglects small corrections proportional to 
the up and charm-quark masses, the resulting terms in the Lagrangian are,
\begin{eqnarray}
 y_{1}~ \left( \phi_{1} c_{R}  - \phi_{2} u_{R}  \right) t_{R}
 +y_{2}~ \phi_{3} u_{R} c_{R} + h.c
\end{eqnarray}
where $u_R$, $c_R$, and $t_R$ are Weyl spinors corresponding to the (right-handed parts of the)
quark mass eigenstates, $y_1$ and $y_2$ are complex dimensionless parameters,
and color indices are implicit (contracted anti symmetrically).  The same corrections from the
top Yukawa can result in large splitting between the masses of $\phi_1$ and
$\phi_2$ (which are themselves expected to be degenerate in the limit where the
up- and charm-quark masses are neglected) and the mass of $\phi_3$.

In summary, when $\phi$ is a color triplet which couples to a pair of up-type quarks,
MFV suggests it is a flavor triplet under $SU(3)_{u_R}$.  The theory is described
by two dimensionless couplings and two masses,
\bea
\left\{ y_1,~~ y_2, ~~ m_{\phi_1},~~ m_{\phi_3} \right\},
\eea
where $m_{\phi_{1}}$ is the (approximately degenerate) masses of the two 
colored scalars which couple
to $u_R t_R$ and $c_R t_R$ with (approximately equal)
coupling $y_1$ and $m_{\phi_3}$ is the mass of the
third scalar with couples to $u_R c_R$ with coupling $y_2$.
\begin{figure}[th]
\subfloat[\label{fig:Annihilation-loop}Annihilation $\chi^{\star}\chi \rightarrow ~{\rm gluons}$ at one loop.]{\includegraphics[scale=0.17]{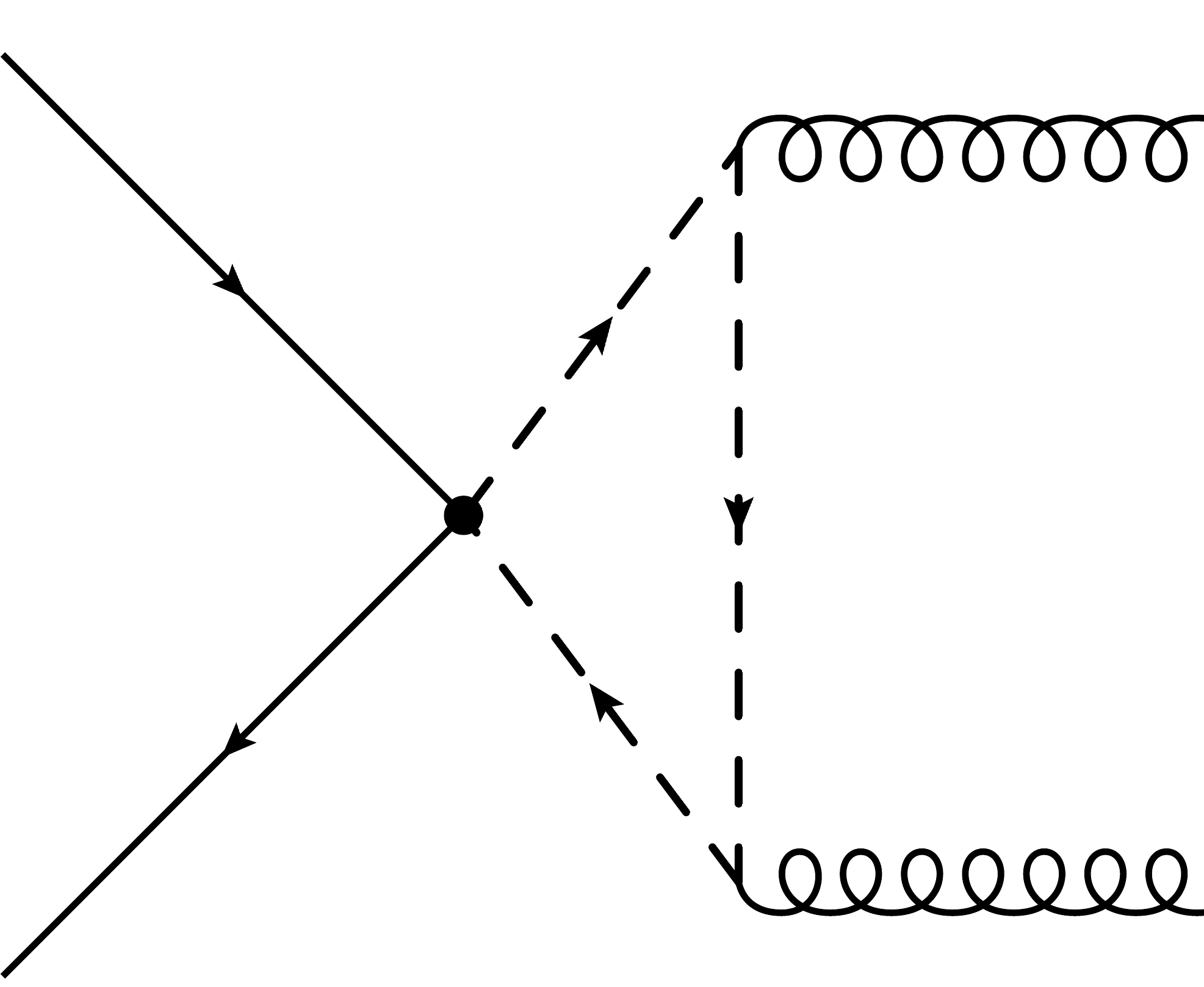}}
\hfill{}
\subfloat[\label{fig:monojet}Mono-jet signature.]{\includegraphics[scale=0.16]{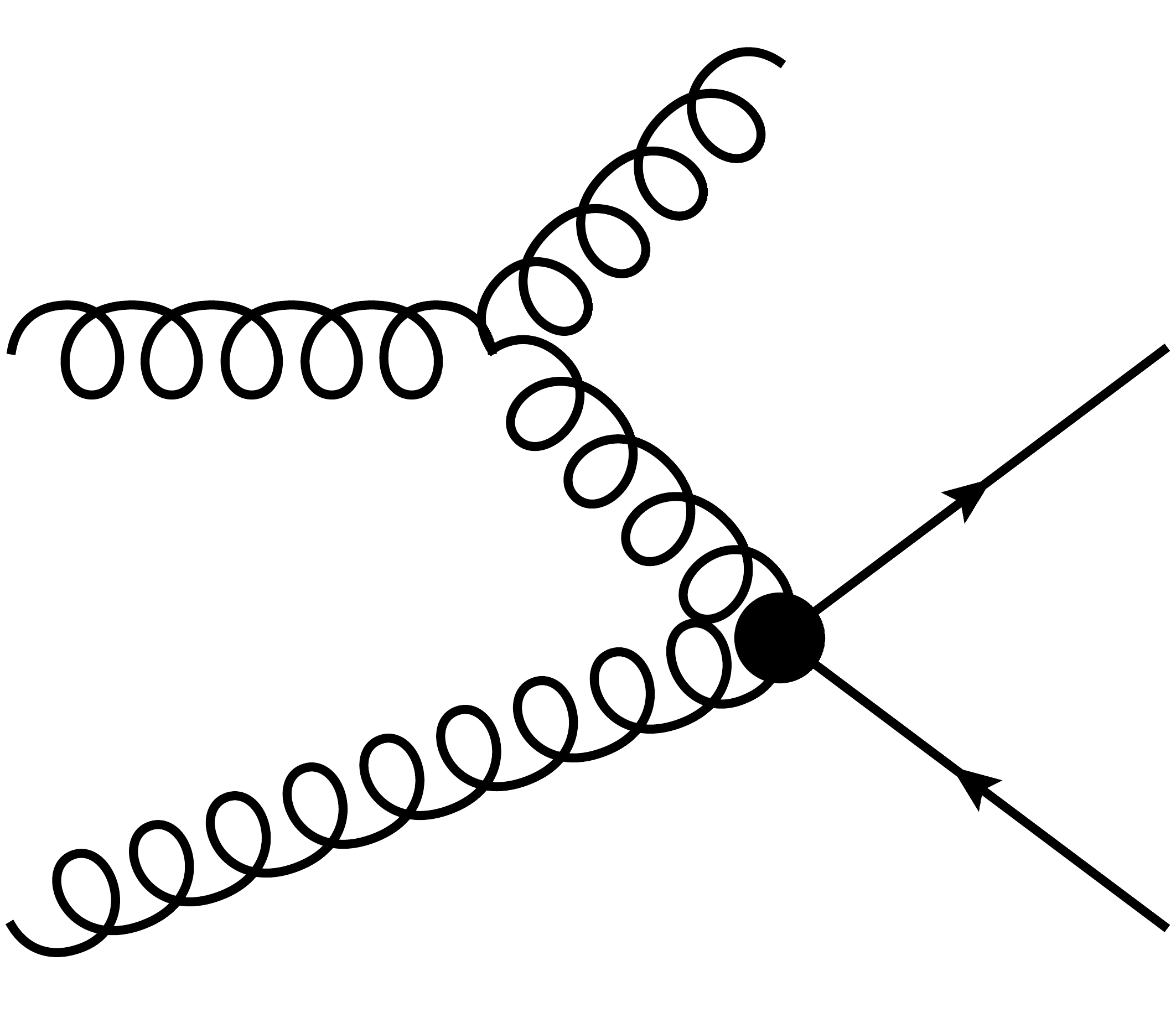}}
\hfill{}
\subfloat[\label{fig:single} Mediator + top quark production followed by decay of the mediator into top and an unflavored jet.]{\includegraphics[scale=0.18]{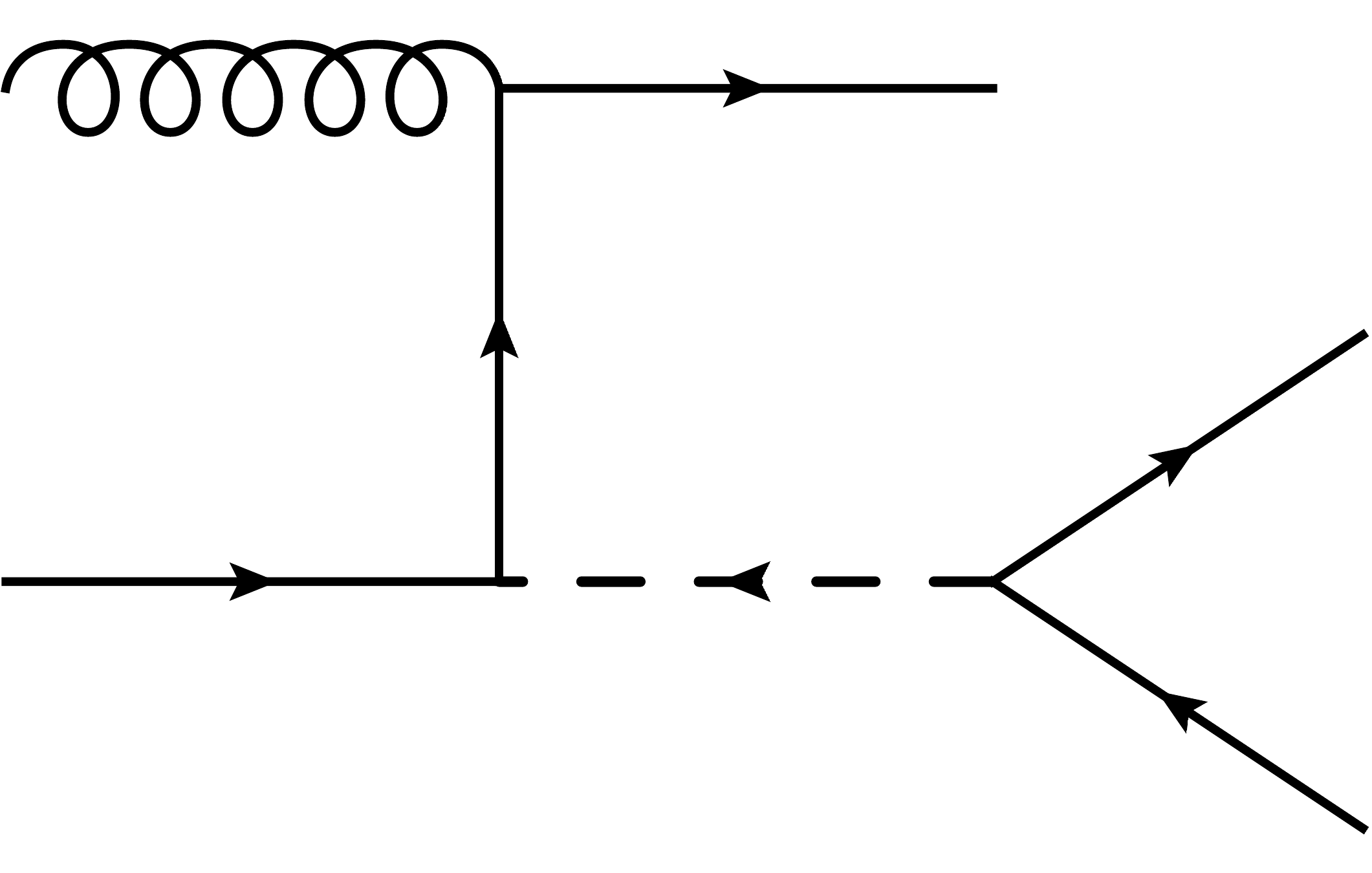}\vspace*{-0.25cm}}
\hfill{}
\subfloat[\label{fig:pair}Pair productoin of mediators followed by decay into two fermions.]{\includegraphics[scale=0.12]{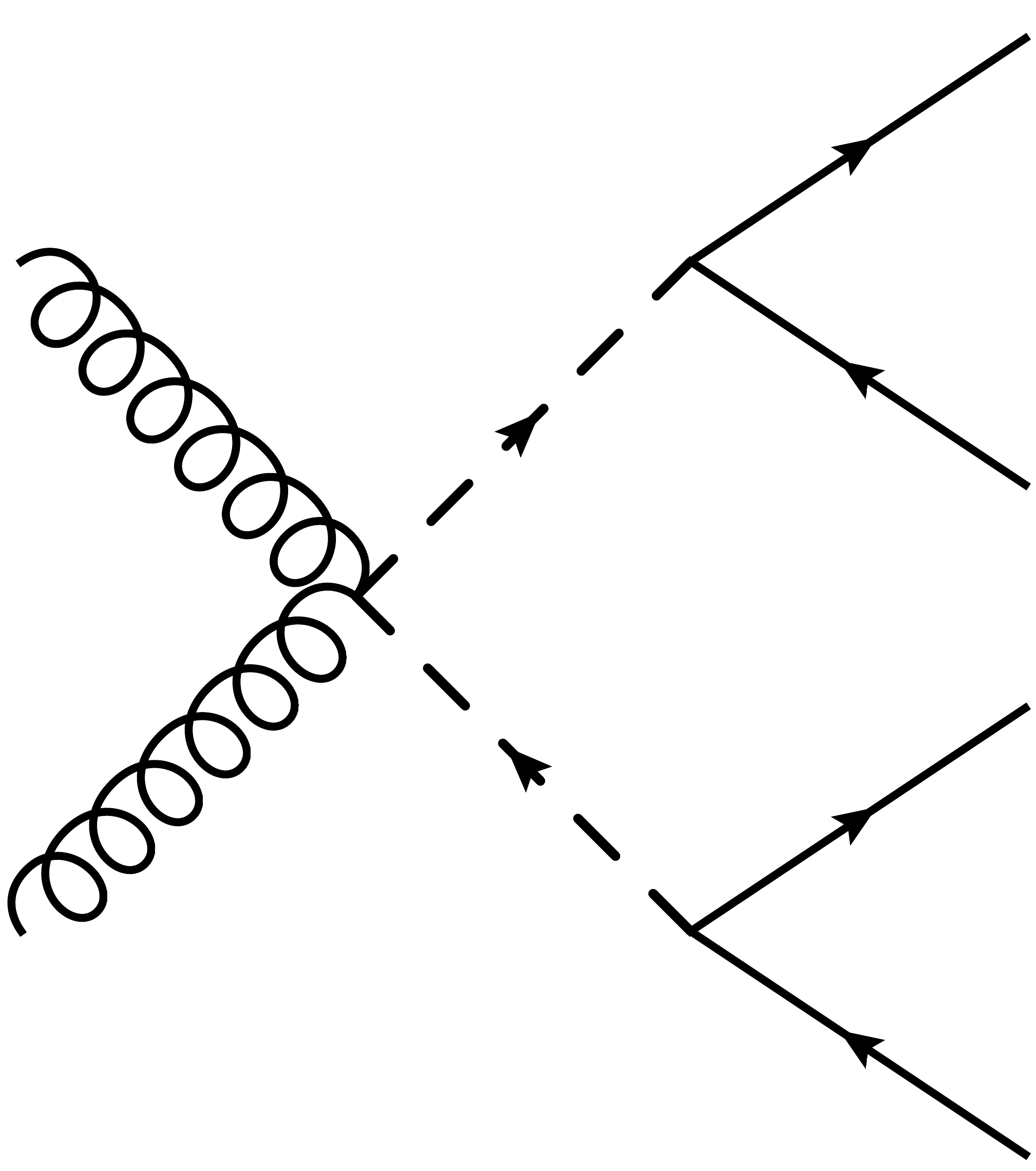} \vspace*{-0.25cm}}
\caption{Representative Feynman diagrams for various processes involving the mediating colored-scalar that we will explore.}
\end{figure}
\section{Annihilation Cross Section\label{sec:Relic-Density}}

The cross section for the dark matter to annihilate is the primary quantity determining the prospects for observing it
via indirect detection methods, and also determining its relic density, if one assumes a standard cosmological $\Lambda$CDM
history.  While we are agnostic toward the actual mechanism responsible for producing $\chi$ in the early Universe,
the relic density singles out a particularly interesting region of parameter space.

The primary mechanism for annihilation depends very sensitively on whether the mediators are heavier or lighter than the
dark matter itself.  When one or more of the mediators are lighter, the annihilation will be dominated by the tree level
process $\chi \chi^* \rightarrow \phi \phi^*$ (where the $\phi$ eventually decay into quarks).  This rate is entirely
controlled by the quartic coupling $\lambda_d$, with rather mild dependence on the mass of $\phi$,
\begin{eqnarray}
\langle \sigma v_{\chi} \rangle & = & \frac{\lambda_d^2 \times r}{64\pi m_{\chi}^{2}} ~\sum_{m_{\phi_i} < m_\chi}~ \sqrt{1-\frac{m_{\phi_i}^{2}}{m_{\chi}^{2}}}
\end{eqnarray}
where $r$ is the color representation of $\phi$ and the sum includes all flavors of $\phi$ whose masses are less than
$m_\chi$.

When all of the mediators have masses larger than $m_\chi$, annihilation can go through off-shell $\phi$'s into quarks
(depending on the strength of the coupling of the $\phi$ particles to the quarks), into Higgs bosons through the
$\lambda_{\chi h}$ coupling, or into gluons through a loop of $\phi$ particles
(see Figure~ \ref{fig:Annihilation-loop}).  The former two processes can be neglected when $\lambda_{\chi h}$ and
the $y_i$ couplings are very small, but the one loop coupling to gluons is also proportional to $\lambda_d$. 
The resultant $\chi \chi^* \rightarrow gg$ cross section can
be expressed in the non-relativistic limit,
\begin{eqnarray}
\langle \sigma v_{\chi} \rangle & = & \frac{\lambda_d^2 ~ T^2_r ~ \alpha_s^2}{64 \pi^{3} m_{\chi}^{2}}
~ \left| \sum_i ~ \left( 1 + 2 m_{\phi_i}^2 C_0 \right) \right|^{2}
\end{eqnarray}
where $C_0$ is the usual scalar three-point integral \cite{Passarino:1978jh} with arguments $C_0(0, 0, 4m_\chi^2; m_{\phi_i}, m_{\phi_i}, m_{\phi_i})$, $T_r$ is the Casimir for representation $r$ of $SU(3)$,
and $\alpha_s \sim 0.1$ is the strong coupling constant.

\begin{figure}[t]
\hspace*{-2cm}
\includegraphics[scale=0.57]{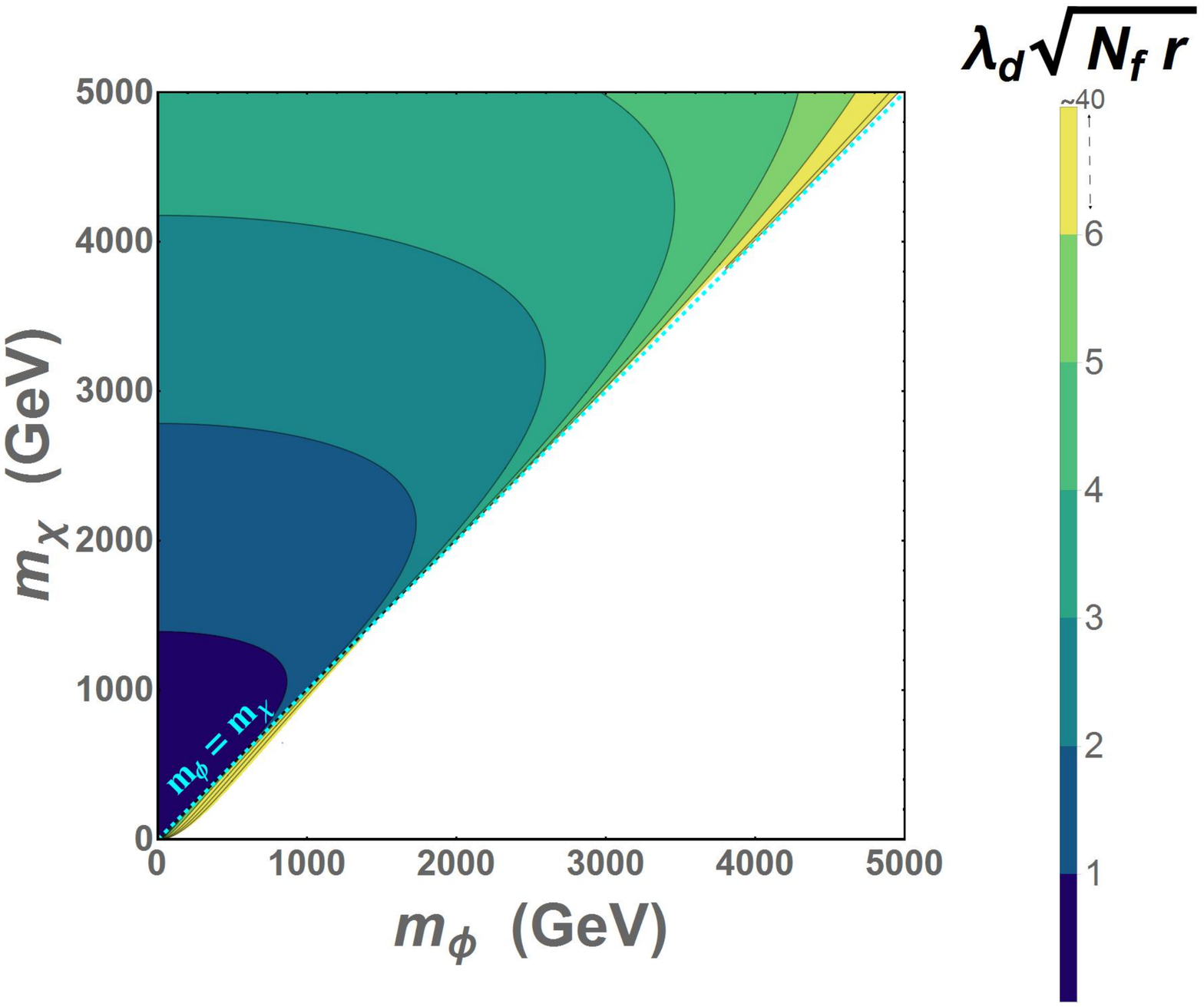}
\caption{\label{fig:Constraints from relic density}
The product of quartic interaction $\lambda_d$ with the square root of product of $r$ dimensional color representation of
$\phi$ and $N_f$ number of flavors with mass less than $m_\chi$, 
required to saturate the observed dark matter density as a thermal relic, are represented as colored contours in the plane of
$m_\phi$-$m_\chi$. Almost all the parameter space where $m_\phi < m_\chi$ is compatible with a thermal relic density. Where $m_\phi > m_\chi$, the DM annihilation proceeds via loops and, only a small region of parameter space is allowed without including any additional couplings.} 
\end{figure}

For a thermal relic in a standard $\Lambda$CDM cosmology, the annihilation cross section should be approximately
$\langle \sigma v \rangle \sim 3 \times 10^{-26} {\rm cm}^{3}{\rm s}^{-1}$ (with a relatively mild dependence
on $m_\chi$ becoming more pronounced for small masses) \cite{Steigman:2012nb}.  Given the loop suppression, it is very difficult to realize
this cross section for all of the 
$m_\phi > m_\chi$ apart from a narrow sliver around $m_\phi \lesssim 1000$~GeV.
In this region, a thermal relic requires one to
invoke large coupling to quarks
$y_i$ or the Higgs
$\lambda_{\chi h}$ to allow the other channels to make up the difference in the cross section.  
For $m_\chi > m_\phi$, one can realize a
thermal relic for a wide variety of masses and couplings $\lambda_d$.
Our results are presented in Figure~ \ref{fig:Constraints from relic density}.

\section{Scattering with Nuclei\label{sec:Direct-detection}}

Direct searches for dark matter are most sensitive to its scattering with heavy nuclei.  The most stringent 
constraints are on spin-independent (SI) 
interactions with nucleons, which are typically coherent for the momentum
transfer of interest to typical experiments ($\sim 10$~MeV) .  
Given the large expectation value of matrix elements for
gluons in the nucleon, the dominant contribution will be
from the effective coupling to gluons induced by loops of the mediators
(c.f. Fig~ \ref{fig:Annihilation-loop}).
\begin{figure}[t]
\hspace*{-0.35cm}
\includegraphics[scale=0.7]{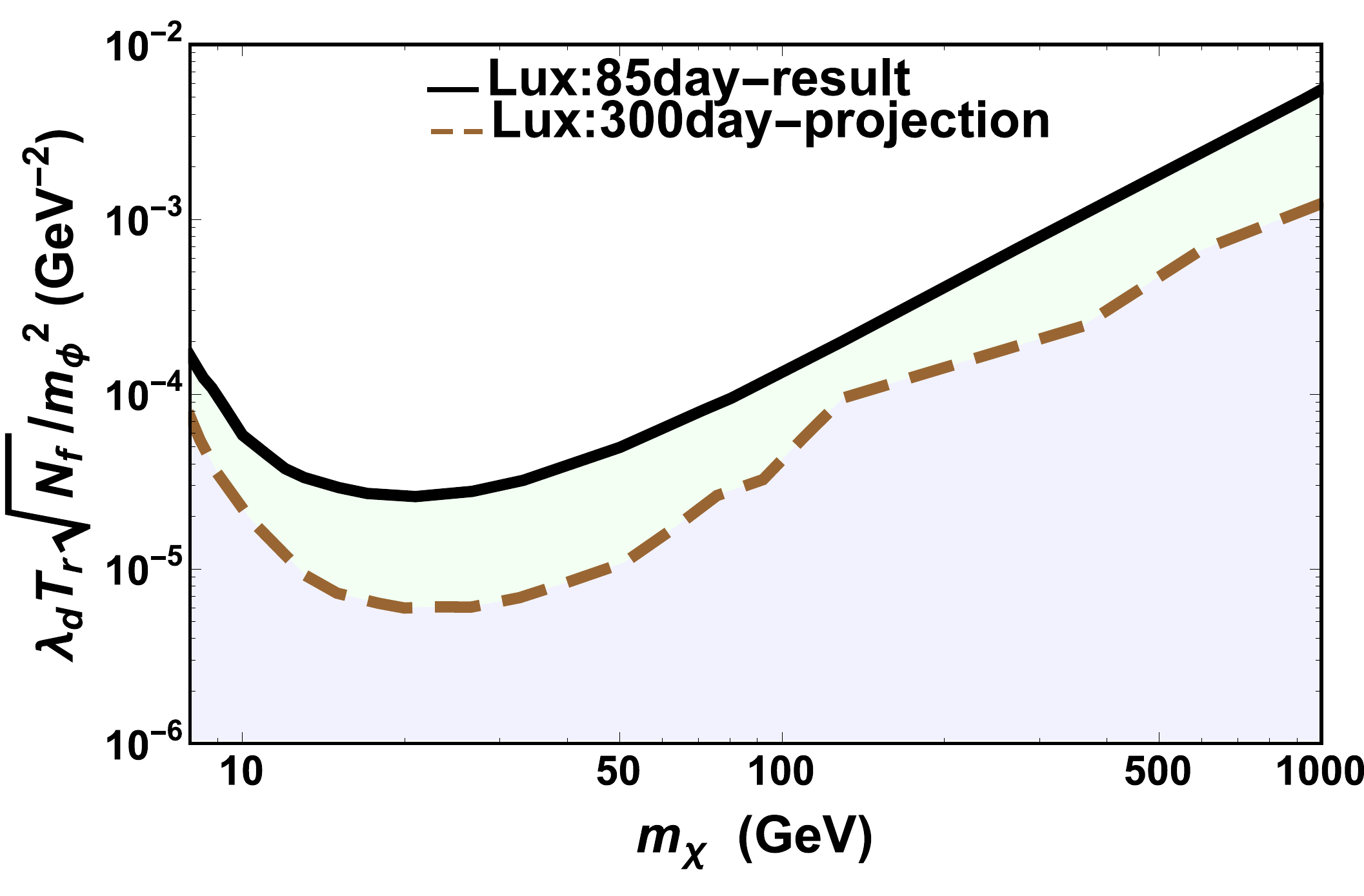}
\caption{\label{fig:Lux bounds}Current (solid line) and projected (dashed line) bounds on 
$\sum \lambda_d T_r \sqrt{N_f} / m_\phi^2$
based on searches for dark matter-Xenon scattering by LUX.  The region above the solid
line is excluded.}
\end{figure}

To good approximation, the coupling to gluons can be represented by its leading term in the expansion
of the momentum transfer divided by the mediator mass.  
In this limit, the effective coupling can be represented by the operator C5,
\bea
\frac{\lambda_d \alpha_s T_r}{48 \pi} \sum_i ~  \frac{1~}{m_{\phi_i}^2} 
~ |\chi|^2 G^a_{\mu \nu} G^{a \mu \nu}~,
\eea
whose coefficient is determined by $\lambda_d$, $T_r$, and the masses of the mediators.  
It is convenient to introduce the masses added in parallel,
\bea
\frac{1}{\overline{m}^2} & \equiv & \sum_i ~  \frac{1~}{m_{\phi_i}^2},
\eea
which in the limit where all mediators have equal masses is $1 / \overline{m}^2 \rightarrow N_f / m_\phi^2$.
Combined with
the gluonic matrix elements, the result is a spin-independent cross section $\sigma_{\rm SI}$,
\bea
5.2 \times 10^{-44}{\rm cm}^2 \left( \lambda_d T_r\right)^2 \hspace*{-0.05cm}
\left( \frac{\mu_\chi~m_\chi}{\rm 10~GeV^2} \right) \hspace*{-0.075cm}
\left( \frac{\rm 200~GeV}{\overline{m}} \right)^4 ,
\eea
where $\mu_\chi$ is the reduced mass of the nucleon - dark matter system.
Through the renormalization group
the gluon operator will mix with the scalar quark bilinear, and is expected to lead to modest
changes to this expectation which grow as the $\log$ of $m_\phi$ \cite{Hill:2014yxa}.

Currently, the most stringent bound on $\sigma_{\rm SI}$ for a wide range of dark matter mass
is obtained from the null observation after
85 days of live running  by the LUX experiment with a liquid Xenon target \cite{Faham:2014hza}.  In 
Figure~ \ref{fig:Lux bounds}, we show the bounds on $ \lambda_d T_r / \overline{m}^2$ as a function
of dark matter mass derived from those bounds, and also compare with projected bounds based on
300 days of live running.  For $\lambda_d T_r \sqrt{N_f} \sim 1$, mediator masses of order 200 GeV remain
consistent with observations.

\section{Collider Constraints\label{sec:Collider-Constraints}}

With an effective coupling to gluons and additional heavy colored states, this simplified model leads
to rich phenomenology at hadron colliders such as the LHC.  Since the mediating scalars do not
themselves decay into the dark matter, the associated phenomenology is quite distinct from the usual
$R$-parity conserving SUSY, with the specific details dependent on the choice of color representation
$r$ as well as the transformation under the quark flavor symmetries of the mediators.

\subsection{Missing Energy Signature}

Independent from the choices of $r$ and flavor embedding, 
the effective coupling to gluons allows production of pairs of dark matter.
When accompanied by additional radiation in the form of quarks or gluons 
(see Figure~ \ref{fig:monojet}), the presence of the
dark matter can be inferred by the imbalance of transverse momentum, resulting in a mono-jet
signature.

In the limit in which the energies of the participating partons are all much smaller 
than $m_\phi$, the dominant interaction can be represented as the contact interaction C5,
much as was the case describing the scattering with nuclei.  For this region of parameter
space, the bounds from ATLAS and CMS  \cite{Khachatryan:2014rra,Aad:2015zva}
apply and (translated from Dirac dark matter to complex scalars)
provide a limit on the same combination of $m_\phi$ and $\lambda_d$.  The most stringent
limit is currently from CMS \cite{Khachatryan:2014rra}, and requires,
\bea
\frac{\lambda_d T_r}{48 \pi} ~  \frac{1~}{\overline{m}^2}
\leq \frac{1}{(207~{\rm GeV})^2}
\eea
for $m_\chi \lesssim 200$~GeV, with the limit weakening to nonexistent as the dark
matter mass approaches $\sim$~1 TeV.

These limits are only reliable for mediator masses well above the typical momentum
transfer in the events, characterized for the CMS analysis by the cut on
missing transverse momentum of $E_{T}^{\rm miss}>  500~{\rm GeV}$.  For small
$r$ and numbers of mediators, there is essentially no useful bound from
the mono-jet search for any perturbative value of $\lambda_d$.  Only 
for very large values of $r$, $N_f$, and/or $\lambda_d$ are there meaningful bounds.

In the regime of smaller $m_{\phi}$, the effective contact interaction softens, driving the signal to
look more like background.  While a detailed study is beyond our scope, it is unlikely that the bounds
are strong enough to be useful in this part of the parameter space. Additionally, bounds obtained in this section are comparable to a detailed analysis of missing energy signature, for an axial vector mediator along with a fermionic dark matter candidate, which obtains a lower bound on the mass of the mediator of the order $\sim1~{\rm TeV}$ for $m_{\chi}\sim~100~{\rm GeV}$ with order 1 couplings and the mediator mass is unconstrained for $m_{\chi}> 300~{\rm GeV}$\cite{Chala:2015ama}.

\subsection{Production of the Colored Mediators}

The mediating scalars interact directly with gluons, and as a result have large pair production cross
sections which to good approximation are functions only of $r$ and the mass of the mediator.  As
outlined above, they typically decay into two or more jets of hadrons, which may favor certain
flavors (depending on the how the weak charges and transformation under the quark flavor
symmetries are chosen).  Such searches are notoriously difficult, with null searches for 
pair production of a pair of color triplet resonances decaying into two-jets
excluding masses below about 350 GeV \cite{Khachatryan:2014lpa}.  For larger $r$ and/or
more complicated transformation under the flavor symmetries, more exotic configurations of
a pair of resonances each decaying to a large multiplicity of jets are possible.  While such
events have spectacular kinematic structure, revealing its existence is complicated by
large combinatoric backgrounds.  We leave the exploration of such novel signatures
for future work.

In the limit in which the coupling to quarks are extremely tiny, the
mediators may be long-lived on collider scales, and the best bounds come from searches
for colored particles stopping in the detector material, and then later decaying ``out of time".
Searches for such objects which are colored triplets or octets result in bounds
on their masses of roughly $m \geq 900$ and $m \geq 1200$~GeV, respectively
\cite{ATLAS:2014fka,Khachatryan:2015jha}.

\subsection{Color Triplet Mediator Coupled to $u_R u_R$}

For our specific example of a color triplet mediator coupled to a pair of
right-handed up-type quarks, there are three mediators, $\phi_{1,2,3}$
which couple to $c_R t_R$, $u_R t_R$, and $u_R, c_R$, respectively.
Since $\phi_3$ decays into unflavored jets of hadrons, the CMS
bound \cite{Khachatryan:2014lpa} on pairs of dijet resonances (see Figure~ \ref{fig:pair})
requires $m_{\phi_3} \geq 350$~GeV.  For large enough $y_2$, there
are bounds coming from searches for resonances in dijet events.  The bounds are
a somewhat complicated collection of searches at different collider energies
(for a nice review, see \cite{Dobrescu:2013cmh}).  The most constraining searches
\cite{Aad:2010ae,Khachatryan:2010jd,Aad:2014aqa,Khachatryan:2015sja} typically require $y_2$ less than about 0.8,
with the most tight constraints ($y \lesssim 0.2$) occurring for masses around 1 TeV
and essentially no constraint on masses greater than about 3 TeV.

\begin{figure}[t]
\includegraphics[scale=0.3]{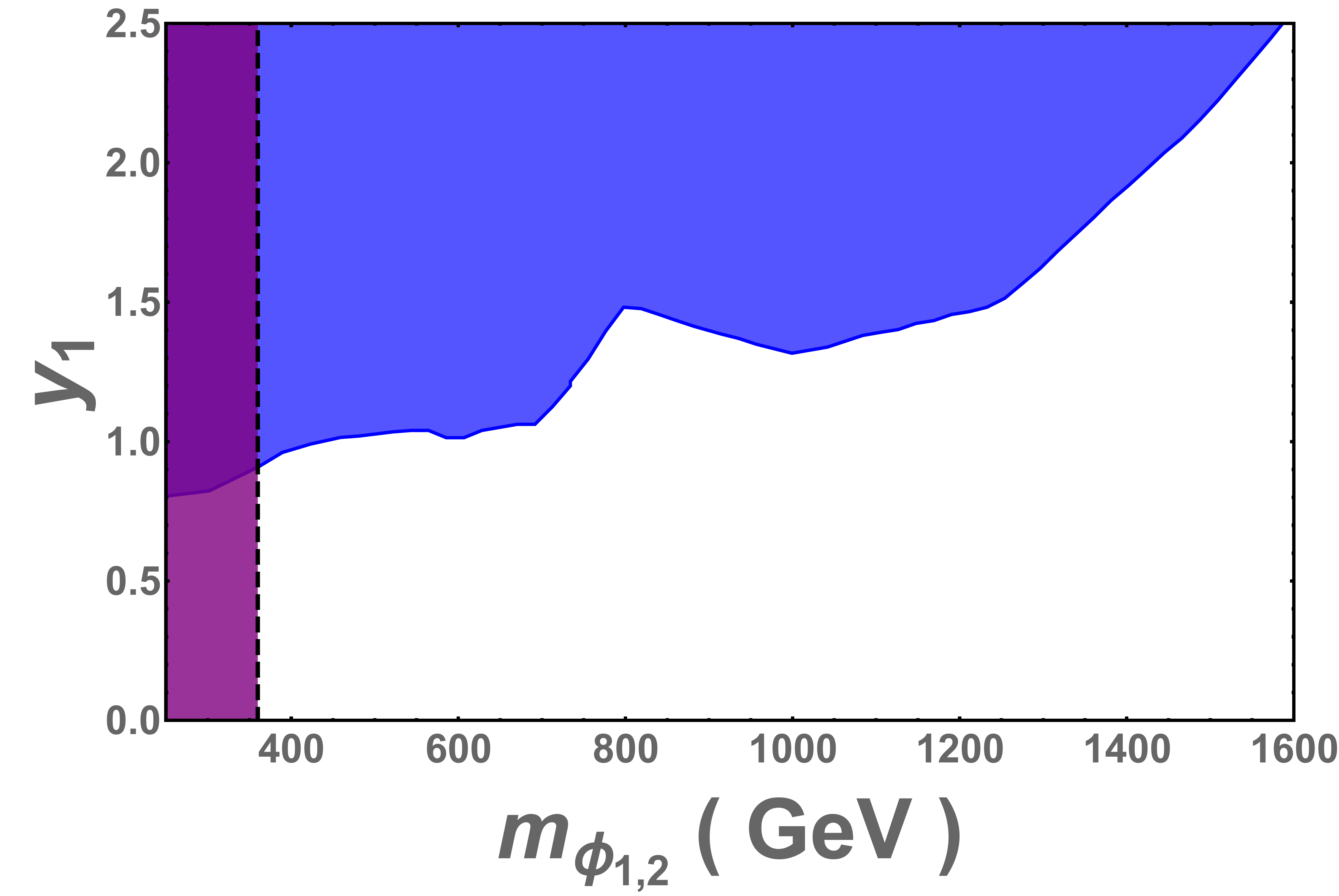}
\caption{\label{fig: triplet collider combined bounds} 
Excluded region of the plane of $m_{\phi_{1,2}}$ and $y_1$ from 
searches for anomalously large production of $t\overline{t} + {\rm one~jet}$
(solid blue region) and $t\overline{t}+ {\rm two~jets}$ (purple shaded region).}
\end{figure}

$\phi_1$ and $\phi_2$ have flavor-changing couplings to the top quark,
together with a charm or up quark, respectively.  Consequently, they
decay into a top and an unflavored jet, resulting in top-rich signatures.  Pair production
of $\phi_{1,2}^* \phi_{1,2}$ thus leads to a $t \bar{t} + 2~{\rm jets}$ signature, with
resonant structure in the invariant masses constructed from 
one of the tops and one of the unflavored jets (Figure~ \ref{fig:pair}).  The rate of this process
is controlled by QCD, and thus depends on the masses of the mediators but not the strength
of their coupling to quarks.  
There is a dedicated CMS analysis for this kinematic structure \cite{CMS:2013gqa}, which for pair
production of (two mass-degenerate) color triplet scalars requires $m_{\phi_{1,2}} \geq 350$~GeV
for essentially any value of the coupling $y_1$ leading to a prompt decay.
The excluded area
is represented as the purple shaded region in Figure~ \ref{fig: triplet collider combined bounds}.

In addition, for small enough masses the contribution
to the inclusive $t \bar{t}$ production could be large enough to disagree with observations.
For two degenerate color triplet scalars ,the inclusive top cross section measurement at 8 TeV 
\cite{Chatrchyan:2012ria,ttbar8TevATLASCMS} provides the weaker
constraint $m_{\phi_{1,2}} \geq 220$~GeV, again roughly independently of the value
of $y_1$.\footnote{We simulate the production cross section at leading order using \texttt{MadGraph
5} \cite{Alwall:2014hca} and include a k-factor of 1.7 extracted
from calculations of squark productions at NLO \cite{GoncalvesNetto:2012yt}.}

If the coupling to quarks is large enough, there is also the possibility of radiating a mediator from
an up or charm quark, resulting in the signature of 
$t \bar{t} +~{\rm unflavored~jet}$ (see Figure~ \ref{fig:single}, with a resonance
between one of the tops and the jet.  There are dedicated searches for this kinematic structure
\cite{Aad:2012em,Chatrchyan:2012su} 
which place bounds in the plane of the masses $m_{\phi_{1,2}}$ and coupling $y_1$.
The resulting excluded region is plotted as the solid area in Figure~ \ref{fig: triplet collider combined bounds}.

\section{Conclusions\label{sec:Conclusions}}

A model in which the dark matter interacts primarily with the Standard Model 
via the gluons (and not appreciably with the quarks) is
an interesting corner of dark matter theory space, one worthy of both theoretical and experimental 
exploration.  We construct an appealing
renormalizable simplified model in which the dark matter is a scalar particle, whose coupling to 
gluons is induced through a quartic
interaction connecting it to exotic colored scalars.  A large number of choices for color and flavor 
representations of the scalars
exist, though all share some common features.  In particular, the strongest 
constraints (for $m_\chi \gtrsim 10$~GeV) 
typically come from direct searches
for dark matter scattering with nuclei, with missing energy signals at the LHC strongly suppressed.  
The colored scalars themselves
typically decay into a number of quarks, motivating searches at the LHC for multi-jet signals 
of resonantly produced pairs
of particles with QCD-sized production cross sections.   

It is perhaps surprising that some models of dark matter may manifest themselves
at a hadron collider most readily through a signature {\em without} any missing transverse momentum.

\acknowledgments

The authors are pleased to acknowledge conversations with Eva Halkiadakis, Arvind Rajaraman, Flip Tanedo, and
Scott Thomas.
The research of TMPT is supported in part by NSF grant PHY-1316792
and by the University of California, Irvine through a Chancellor's Fellowship. The research of GM was supported by CSIR, India via SPM fellowship-07/079(0095)/2011-EMR-I. GM would also like to thank the particle physics community at the Department of Physics and Astronomy  at University of California,  Irvine, USA for their kind hospitality from Mar'14-May'14 and Sept'14-Jan'15 where a part of this work was accomplished. RMG wishes to acknowledge support from the Department of Science and Technology, India under Grant  No. SR/S2/JCB-64/2007 , under the J.C. Bose Fellowship scheme.
\bibliographystyle{JHEP}
\bibliography{workpapers}

\end{document}